\documentclass[conference]{IEEEtran}
\ifCLASSINFOpdf
\else
\usepackage{graphicx}
\graphicspath{{../eps/}}
\DeclareGraphicsExtensions{.eps}
\fi

\usepackage{amsmath}
\usepackage{amsfonts,amssymb}
\usepackage{amssymb}
\usepackage{wrapfig}
\usepackage{psfrag}
\usepackage{epstopdf}
\usepackage{cite}
\usepackage{graphicx}
\usepackage{subfigure}
\usepackage{threeparttable}
\usepackage{cases}
\usepackage{subeqnarray}
\usepackage{color}
\usepackage{underscore}
\usepackage{verbatim}
\usepackage{bm}
\usepackage{stfloats}

\usepackage{algorithm}
\usepackage{algorithmic}

\begin{document}
	
	% paper title
	\title{Environment-Aware and Training-Free Beam Alignment for mmWave Massive MIMO via Channel Knowledge Map}
	%
	%
	%%
	%%
	%% author names and IEEE memberships
	%% note positions of commas and nonbreaking spaces ( ~ ) LaTeX will not break
	%% a structure at a ~ so this keeps an author's name from being broken across
	%% two lines.
	%% use \thanks{} to gain access to the first footnote area
	%% a separate \thanks must be used for each paragraph as LaTeX2e's \thanks
	%% was not built to handle multiple paragraphs
	%%
	%
	
	\author{\IEEEauthorblockN{$\text{Di~Wu}^*$, $\text{Yong~Zeng}^*\ddagger$, $\text{Shi~Jin}^*$, and $\text{Rui~Zhang}^\dagger$ }
		\IEEEauthorblockA{*National Mobile Communications Research Laboratory, Southeast University, Nanjing 210096, China\\
			$\ddagger$Purple Mountain Laboratories, Nanjing 211111, China\\
			$\dagger$Department of Electrical and Computer Engineering, National University of Singapore, Singapore 117583\\
			220200691@seu.edu.cn, yong_zeng@seu.edu.cn, jinshi@seu.edu.cn, elezhang@nus.edu.sg}
	}
	
	% make the title area
	\maketitle

\begin{abstract}
Millimeter wave (mmWave) massive multiple-input multiple-output (MIMO) communication system is expected to achieve enormous transmission rate, provided that the transmit and receive beams are properly aligned with the MIMO channel. However, existing beam alignment techniques rely on either channel estimation or beam sweeping, which incur prohibitively high training overhead, especially for future wireless systems with further increased antenna dimensions and more stringent requirement on cost-effective hardware architectures. 
In this paper, we propose a new beam alignment technique, which is environment-aware and training-free, by utilizing the emerging concept of channel knowledge map (CKM), together with the user location information that is readily available in contemporary wireless systems. CKM is a site-specific database, tagged with the transmitter/receiver locations, which contains useful channel information to facilitate or even obviate real-time channel state information (CSI) acquistion. Two instances of CKM are proposed for beam alignment in mmWave massive MIMO systems, namely \emph{channel path map} (CPM) and \emph{beam index map} (BIM). 
It is shown that compared with existing training-based beam alignment schemes, the proposed CKM-enabled environment-aware beam alignment is able to drastically improve the effective communication rate, even with moderate user location errors, thanks to its significant saving of the prohibitive training overhead. 
\end{abstract}

%\begin{keywords}
%channel knowledge map, training-free communication, channel path map, beam index map
%\end{keywords}

\IEEEpeerreviewmaketitle
\section{Introduction}
Millimeter wave (mmWave) massive multiple-input multiple-output (MIMO) communication systems are able to achieve enormous transmission rate, thanks to the combined benefits of the wide bandwidth of mmWave spectrum and the high beamforming gain of massive MIMO\cite{MIMO,overview}.
However, compared to the predominating sub-6GHz wireless system nowadays, mmWave signal suffers from much higher free-space attenuation and penetration loss due to its smaller wavelength.
Therefore, in order to achieve comparable spectral efficiency and coverage range for future wireless systems, antenna arrays of much more elements need to be equipped at the base station (BS) than today's standard. Besides, the user equipment (UE) also needs to be equipped with a moderate antenna array  so as to achieve a combined transmit and receive beamforming gain.
To practically reap such beamforming gains, the BS and UE need to align their beams with the MIMO channel properly. 
While beamforming optimization has been well understood in theory for fully digital MIMO systems when perfect channel state information (CSI) is available, its practical implementation for mmWave massive MIMO systems with the cost-effective analog beamforming or hybrid analog/digital beamforming architectures is still challenging \cite{millimeter,variable,phase,analog,hybrid}. This is mainly due to the limited signal processing with much fewer radio frequency (RF) chains than antennas, together with the necessity to acquire CSI in real time. 

There are two major approaches in the literature for mmWave beam alignment, namely {\it training-based CSI estimation} \cite{spatially,estimate} and {\it training-based beam sweeping} \cite{beamsweeping,tutora}. In the first approach, the full MIMO channel matrix is firstly estimated by sending pilot signals, based on which the transmit and receive beamforming vectors are then designed.
However, different from fully digital MIMO systems \cite{howmu}, channel estimation for mmWave massive MIMO communication with analog or hybrid beamforming generally requires prohibitively high training overhead. This is because on one hand, an excessively large number of channel coefficients need to be estimated, and on the other hand, the channel measured in digital baseband is intertwined with the analog beamforming vectors used to train the channel and the actual MIMO channel matrix cannot be directly accessed. 
While such issues can be relieved to certain extent by techniques such as  compressive sensing \cite{cs}, they rely on additional assumptions on e.g. channel spatial sparsity and perfect knowledge of array geometry and orientations, and also require high computational complexity due to the sophisticated iterative algorithms for signal processing.
In contrast, for the training-based beam sweeping based approach \cite{tutora}, there is no need to estimate the MIMO channel explicitly. Instead, the transmit and receive beams from predefined codebooks are sequentially applied with handshaking between transmitter and receiver, based on which the best beam pairs for them are selected. While beam sweeping avoids the sophisticated CSI estimation and beam selection, it still incurs high training overhead due to the exhaustive search over both transmit and receive beamforming codebooks. Various attempts have been made to reduce the training overhead for beam sweeping, such as using hierarchical multi-resolution codebook \cite{hcodebook}; however, the training overhead is still practically high, especially for future wireless systems increasing number of antennas, such as the extremely large-scale MIMO system \cite{luhaiquan}.  

In this paper, we propose a novel approach for beam alignment, which is environment-aware and training-free, by utilizing the emerging concept of channel knowledge map (CKM) \cite{CKM}, together with the user location information that is readily available in contemporary wireless systems. Note that utilizing various geolocation-based database in wireless communication systems has received growing attention recently \cite{R1, R2, R3}. In particular, CKM aims to enable radio environment-awareness by offering whatever relevant information regarding the intrinsic propagation channels, so as to facilitate or even avoid sophisticated real-time CSI acquisition \cite{CKM}. Some simple examples of CKM are channel gain map (CGM) \cite{Channel4C} and channel shadowing map (CSM). In this paper, we propose two novel instances of CKM for environment-aware and training-free beam alignment in mmWave massive MIMO systems, namely {\it channel path map} (CPM) and {\it beam index map} (BIM). Specifically, CPM aims to predict the channel path information for all potential UE locations in the coverage area, such as the number of significant channel paths and their amplitudes, phases, angle of arrivals (AoAs) and angle of departures (AoDs). With CPM, together with the UE location information, the key parameters of the MIMO channels can be attained and the MIMO channel matrix can be reconstructed without relying on traditional channel training. 
On the other hand, with BIM, the MIMO channel does not need to be estimated explicitly, while instead, the indices of the best transmit and receive beam pairs are learned for all potential UE locations.
Extensive simulations are performed based on realistic communication environment with the aid of commercial ray-tracing software.
The simulation results demonstrate that compared to the conventional training-based methods, the proposed CKM-enabled environment-aware beam alignment schemes can significantly improve the effective communication rate even with moderate UE location errors, thanks to its significant saving of training overhead.

\section{System Model}

As shown in the Fig. 1, we consider a downlink mmWave massive MIMO communication system, which consists of a BS with $ M_t \gg 1$ transmit antennas and a UE with $ M_r>1 $ receive antennas\footnote{The proposed technique can be straightforwardly applied to the uplink communication and/or serving multiple UEs via time division multiple access (TDMA).}. 
For ease of exposition, the BS and each UE are assumed to have one RF chain, and beamforming is performed in the analog domain via phase shifters. Denote the beamforming vectors at the BS and the UE of interest as $ \mathbf{f}\in \mathbb{C}^{M_t \times 1} $ and $ \mathbf{w}\in \mathbb{C}^{M_r \times 1} $, respectively. 
With analog beamforming, each element of $ \mathbf{f} $ and $ \mathbf{w} $ has constant modulus, i.e., $  |\mathbf{[f]}_i|=1/\sqrt{M_t}\ \text{and}\ |\mathbf{[w]}_j|=1/\sqrt{M_r},\ i=1,\ldots,M_t, \  j=1,\ldots,M_r$.
Furthermore, we assume codebook-based analog beamforming, where $\mathbf{f}$ and $\mathbf{w}$ are selected from the pre-determined codebooks
$ \mathcal{F} $ and $ \mathcal{W} $ at the BS and UE sides, respectively. The total number of candidate beamforming vectors are denoted by $ |\mathcal{F}| $ and $|\mathcal{W}|$, respectively. Usually, we have $ |\mathcal{F}|\geq M_t $ and $ |\mathcal{W}|\geq M_r $. 
\begin{figure}[htbp]
 \centering{\includegraphics[width=.50\textwidth]{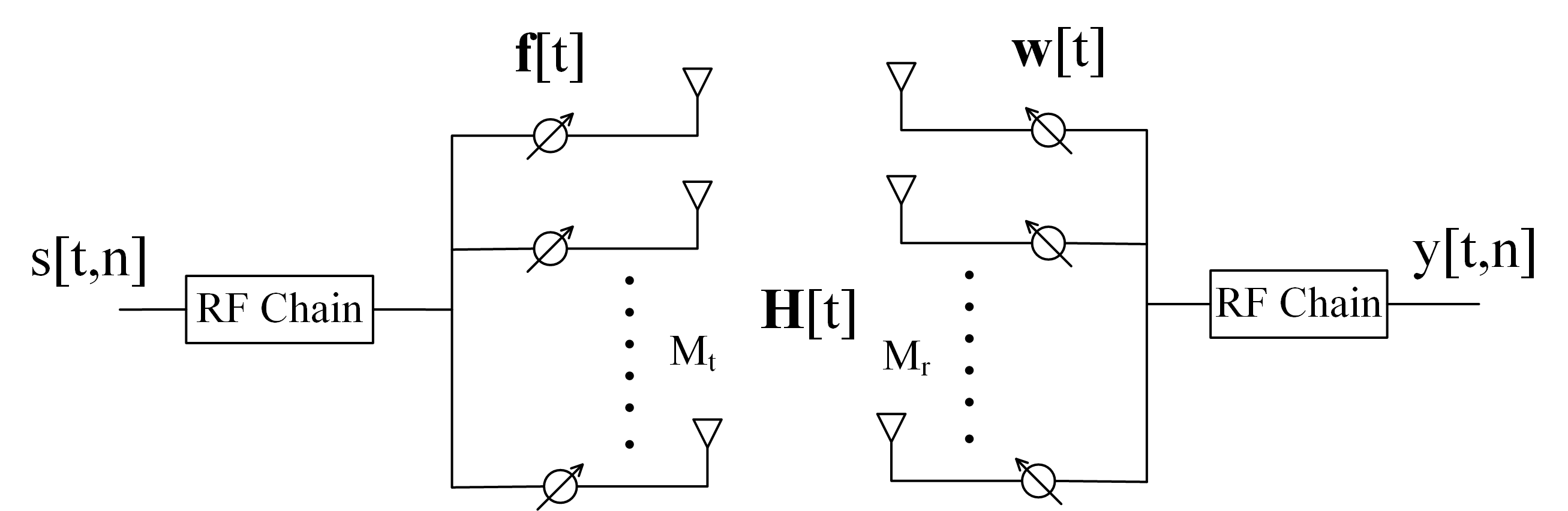}}  	
	\caption{MmWave massive MIMO communication with analog  beamforming.} \label{fig}  
\end{figure}

With quasi-static block fading channel model, the MIMO channel coefficients remain constant for each channel coherent block of $ N $ symbol durations, and may vary across different blocks.
%	When Tx transmits data stream $ s[n] $ (n is the symbol index) with unit-energy $ \mathbf{E}\{|s[n]|^2\}=1 $ to the Rx, the output signal of the combiner at the Rx is
For each coherent block $ t $, denote the channel matrix as $ \mathbf{H}[t]\in \mathbb C^{M_r\times M_t} $. For the MIMO system over $ T $ channel coherent blocks, the received signal at the UE is given by 
\begin{equation}
	\begin{aligned}
			y[t,n]=\sqrt{P}\mathbf{w}^{H}[t] \mathbf{H}[t]\mathbf{f}[t] s[t,n]+\mathbf{w}[t]^{H}\mathbf{n}[t,n], \quad \\
		n=1,2,\ldots,N,\ t=1,2,\ldots, T,
	\end{aligned}
\end{equation}
where $P$ is the BS transmit power that is assumed to be a constant, $ s[t,n] $ denotes the $ n $th information-bearing symbol of coherent block $ t $ with normalized power, i.e., $ E[|s[t,n]|^2]=1 $, and $\mathbf{n}[t,n]\in\mathbb{C}^{M_\mathrm{r}\times 1}$ denotes the zero-mean circularly symmetric complex Gaussian (CSCG) noise vector with variance $  \sigma^2$, i.e., $ \mathbf{n}[t,n] \sim $  $\mathcal{CN}(0,\sigma^2\mathbf{I}) $. Furthermore, $ \mathbf{f}[t]\in \mathcal{F} $ and $\mathbf{w}[t]\in \mathcal{W}$ denote the selected transmit and receive beamforming vectors during coherent block $t$ at the BS and UE, respectively. For the time horizon of $T$ coherent blocks, if perfect CSI $ \{\mathbf{H}[t]\}_{t=1}^T $ is available and assuming $ N $ is sufficiently large, the maximum average communication rate in bits per second per hertz (bps/Hz) is given by
\begin{equation}
	R^{*}=\frac{1}{T} \sum_{t=1}^{T} \log _{2}\left(1+\frac{P \left| \mathbf{w}_\text{opt}^{H} [t] \mathbf{H}[t] \mathbf{f}_\text{opt}[t]\right|^2}{\sigma^{2}}\right),\ t=1,...,T,
\end{equation}
where $  (\mathbf{w}_\text{opt}[t], \mathbf{f}_\text{opt}[t]) $ is the optimal beamforming pairs from the codebooks $\mathcal{F}$ and $\mathcal{W}$ at channel coherent block $ t $, i.e., 
\begin{equation}
	\begin{aligned}
		\left(\mathbf{w}_{\text {opt}}[t], \mathbf{f}_{\text {opt }}[t]\right)=\arg \max _{\mathbf{f} \in \mathcal{F}, \mathbf{w} \in \mathcal{W}}\left|\mathbf{w}^{H} \mathbf{H}[t] \mathbf{f}\right|^{2},\ t=1,...,T.
	\end{aligned}
\end{equation}

However, in practice, achieving the maximum communication rate in (2) with the optimal beamforming in (3) faces critical challenges. In particular, the channel matrix $ \mathbf{H}[t] \in \mathbb{C}^{M_r \times M_t} $ for each coherent block $ t $ needs to be practically estimated, which incurs prohibitively high training overhead for large MIMO systems with only limited RF chains. Specifically, for each coherent block $ t $, let $N_{tr}\leq N$ be the number of symbol durations used for channel training. With analog beamforming, to estimate the MIMO channel matrix $ \mathbf{H}[t]  $ that has $ M_r M_t $ elements, $ N_{tr}\geq M_rM_t $ is required. Let $\hat{\mathbf{H}}[t]$ denote the estimated channel matrix at coherent block $t$, based on which the transmit and receive beam pairs $\hat{\mathbf{f}}[t]$ and $\hat{\mathbf{w}}[t]$ are selected, i.e., 
\begin{equation}
	\begin{aligned}
		(\hat{\mathbf{w}}[t], \hat{\mathbf{f}}[t])=\arg \max _{\mathbf{f} \in \mathcal{F}, \mathbf{w} \in \mathcal{W}}\left|\mathbf{w}^{H} \hat{\mathbf{H}}[t] \mathbf{f}\right|^{2} .
	\end{aligned}
\end{equation}

As a result, with channel training-based beam alignment, the effective average communication rate is
\begin{equation}
	\begin{aligned}
		R_{t r} &=\frac{1}{T} \sum_{t=1}^{T} \frac{N-N_{t r}}{N} \log _{2}\left(1+\frac{P\left|\hat{\mathbf{w}}^{H}[t] \mathbf{H}[t] \hat{\mathbf{f}}[t]\right|^{2}}{\sigma^{2}}\right) \\
		& \leq\left(1-\frac{M_{r} M_{t}}{N}\right) \frac{1}{T} \sum_{t=1}^{T} \log _{2}\left(1+\frac{P|\hat{\mathbf{w}}^{H}[t] \mathbf{H}[t] \hat{\mathbf{f}}[t]|^{2}}{\sigma^{2}}\right) \\
		& \leq\left(1-\frac{M_{r} M_{t}}{N}\right) R^{*},
	\end{aligned}
\end{equation}
where the pre-log factor $ (N-N_{tr})/N $ accounts for the training overhead; the first inequality follows since $ N_{tr}\geq M_rM_t $; and the second inequality is due to the fact that  $  (\hat{\mathbf{w}}[t], \hat{\mathbf{f}}[t]) $ is selected based on the estimated channel matrix $ \hat{\mathbf{H}}[t] $, which is in general different from the true channel $ \mathbf{H}[t] $. 
The result in (5) shows that for large-scale MIMO systems with $ M_rM_t $ comparable to the channel coherent duration $ N $, there is a huge significant rate degradation of the training-based beam alignment from the maximum rate under perfect CSI, $ R^* $. 

Another critical challenge for channel-training based beam alignment lies in the prohibitive computation overhead in (4). For each coherent block, an exhaustive search over all the possible $|\mathcal{F}||\mathcal{W}|\geq M_r M_t$ beam pairs is required, which incurs prohibitive complexity as $M_rM_t\gg 1$. While such a computation overhead problem could be resolved by the beam sweeping technique without explicit channel estimation \cite{tutora}, a rate performance gap similar to (5) still exists due to the training overhead for beam sweeping. 

To reduce the above training/computation overhead, in this paper, we propose a radically new approach, termed {\it training-free beam alignment}, which aims to obviate the costly training for either channel estimation or beam sweeping. The proposed technique utilizes two important information that critically impacts the actual communication channels: namely the UE location, which is readily available in contemporary wireless systems with improving accuracy, and CKM, which is a site-specific database that contains useful channel-related information for all potential UE locations in the site of interest and reflects the actual radio propagation environment.

\section{Training-Free Beam Alignment Via Channel Knowledge Map}
\subsection{A Fresh New Look at Wireless Channels}
%In the mmWave MIMO system, due to the increase in the number of antennas, the training based schemes suffer from large training overhead. 
%As previously stated, the wireless communication channel is mainly determined by the location of BS and UE as well as the environment of signal transmission. 
%Thus, the accurate UE location and perceived signal propagation environment could be used to reduce training overhead.
Before presenting the details of our proposed training-free beam alignment technique, it is necessary to take a fresh new look at wireless communication channels. At an abstract level, a wireless channel is mainly determined by the radio wave property (such as wavelength), the locations of the transmitter and receiver, and the actual radio propagation environment. Over the past few decades, numerous efforts have been devoted to the mathematical characterization of wireless channels using stochastic and/or geometric-based approaches \cite{doc}. However, existing channel modeling approaches only utilize the partial information  of the transmitter/receiver locations (such as the link distance only) and the very coarse environment information (such as urban, suburban, rural area only). While such modeling approaches are tractable and effective for generalization, their modeled channels inevitably incur non-negligible errors when applied in actual communication scenarios\cite{CKM}, which thus necessitates the real-time channel estimation via pilot-based channel training. On the other hand, with the continuous advancement of localization technologies and enhanced environment awareness, there is  a paradigm shift for environment-aware and training-free wireless communications to resolve the issue of prohibitive training overhead in large-dimension MIMO systems \cite{CKM}. 

%We assume that the BS location remains unchanged, so the channel matrix in the channel coherent time $ t $ within environment coherent time $ T $ can be described as the function of UE location $ q[t] $ at and propagation environment $ E[T] $  i.e. $ H(q[t],E[T]) $. 

For the considered mmWave massive MIMO communication system, with the BS location being fixed, the variation of channel matrix $  \mathbf{H}[t] $ is mainly due to the change of UE locations, which are denoted as $\mathbf {q}[t]$, and/or the variation of the actual communication environment, which is denoted as $E[t]$ for brevity. Therefore,  an abstract representation of the channel matrix $ \mathbf{H}[t] $ can be written as 
\begin{equation}
	\mathbf{H}[t]=g_1(\mathbf{q}[t], E[t]),  \ t=1,2,....,T,
\end{equation}
where $g_1(.,.)$ is an unknown function. In fact, though various approximations are possible, such as that based on stochastic models, it may not be possible to obtain accurate expressions for $g_1(.,.)$ in complex environment, due to the difficulty in mathematically modeling the environment $ E[t] $ accurately, as well as the sophisticated interaction between radio waves and actual environment. Fortunately, such an issue can be circumvented based on the recently proposed concept of CKM \cite{CKM}, which is a site-specific database, tagged with the locations of the transmitters and/or receivers and containing channel-related information to enable environment-awareness, as illustrated in Fig. 2.

One straightforward example of CKM is the channel matrix map (CMM), which directly provides the channel coefficient matrix $\hat {\mathbf H}[t]$ for any given UE location $\mathbf q[t]$. 
Specifically, let $\mathcal Q$ denote the location space that includes all potential UE locations within the BS coverage, i.e., $\mathbf q[t]\in \mathcal Q$, $\forall t$. 
Then a CMM, which is denoted as $\mathcal {M}_{CMM}[t]$, is a mapping $\mathbf q[t]\in \mathcal Q \rightarrow \hat {\mathbf H}$. 
The map $\mathcal{M}_{CMM}[t]$ is time-dependent in general, accounting for the time variation of radio propagation environment.
However, in most practical scenarios, the wireless propagation environment (such as the locations, heights, and dielectric properties of surrounding objects) changes in a much larger time scale than the UE locations, as illustrated in Fig. 2. 
Furthermore, even for those environment factors (such as pedestrians) that may vary with comparable time scale as UE locations, the impact of the former on wireless channels is much less than the latter in practice.
As a result, the update of the CMM or CKM in general ,which is necessary only when there is significant environment change (that can be monitored by dedicated environment-sensing nodes as shown in Fig. 2), would require a much larger time scale than the channel coherent time. 
As a result, provided that sufficiently accurate UE locations $\{\mathbf q[t]\}_{t=1}^T$ are available (say via GPS, Beidou, and other advanced localization technologies), the channel matrix $  \hat {\mathbf{H}}[t] $ can be approximately obtained with the CMM $\mathcal {M}_{CMM}[t]$, without any channel training needed .

However, a CMM that aims to directly attain the MIMO channel coefficients requires excessive computation and storage resources. To address this issue, in the following, we propose two other promising instances of CKM for beam alignment, namely CPM and BIM.         
\begin{figure}[htbp] 				\centering{\includegraphics[width=.5\textwidth]{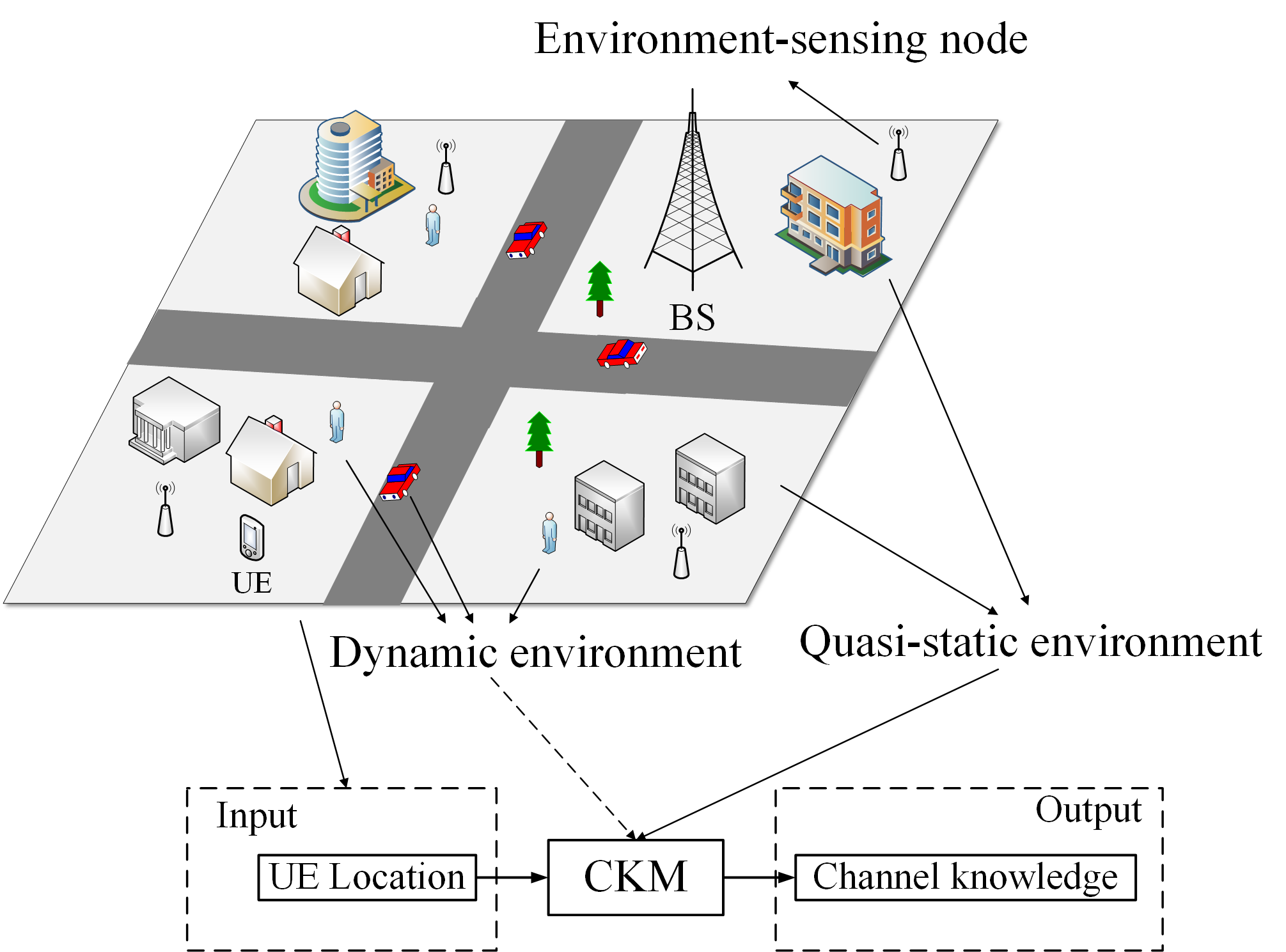}}  	 		 		\caption{An illustration of CKM for enabling environment-awareness. } \label{fig} 
	 	 		 	
\end{figure}
%An example of the CKM-based beam alignment is shown in Fig.3. The BS and UE need to select appropriate beams from their own codebooks to perform beam alignment. Based on the CKM, the BS and UE automatically avoid the linear link blocked by the building, make full use of other surrounding environments, and select the optimal beams to achieve environment-aware communication. Two specific CKM, i.e. CPM and BIM, and beam alignment schemes based on them are illustrated in the following.

%\begin{figure}[htbp] \centering{\includegraphics[width=.5\textwidth]{fig4.png}}  	
%	\caption{An example of CKM-based beam alignment.} \label{fig}  	
%\end{figure}
\subsection{CPM Enabled Training-Free Beam Alignment}

CPM aims to offer location-specific channel path information (such as the number of significant paths and their power, phases, and AoAs/AoDs) for all potential UE locations. Compared with CMM, CPM could be utilized to reconstruct the MIMO channel matrix based on the essential path information with much reduced storage and computation requirements. This is motivated by the classical geometry-based channel model, but with location-dependent path parameters, i.e.,   
\begin{equation}
	\begin{aligned}			 
		\mathbf{H}[t]=\sqrt{M_{\mathrm{r}} M_{\mathrm{t}}} \sum^{L[t]}_{l=1} \alpha_{l}[t]
		\mathbf{a}_{\mathrm{r}}(\theta_\mathrm{AoA}^l[t],\phi_\mathrm{AoA}^l[t]) \mathbf{a}_{\mathrm{t}}^H(\\\theta_\mathrm{AoD}^l[t],
		\phi_\mathrm{AoD}^l[t]),\ t=1,...,T,
	\end{aligned}	 	
\end{equation}
where $ L[t] $ is the number of significant channel paths at channel coherent block $ t $, $ \alpha_{l}[t] $ is the complex gain of the $ l^{th} $ path,  $ \theta_\mathrm{AoA}^l[t] $ and $ \phi_\mathrm{AoA}^l[t] $ are the zenith and azimuth AoA,  $ \theta_\mathrm{AoD}^l[t] $ and $ \phi_\mathrm{AoD}^l[t] $ are the zenith and azimuth AoD of the $ l^{th} $ path, $ \mathbf{a}_{\mathrm{t}}(.,.) $ and $ \mathbf{a}_{\mathrm{r}}(.,.) $ represent the transmit and receive array response vectors, respectively.

At each channel coherent block $t$, denote the channel path information as $ \mathbf{z}[t] $, which includes the number of paths and their power, phases, and AoAs/AoDs, i.e., 
\begin{equation}
	\begin{aligned}			 
	\mathbf{z}[t]=\{L[t], \{ \alpha_l[t], \theta_\mathrm{AoD}^l[t],\phi_\mathrm{AoD}^l[t], \theta_\mathrm{AoA}^l[t],\\ \phi_\mathrm{AoA}^l[t]\}_{l=1}^{L[t]}\}, t=1,...,T.
	\end{aligned}	 	
\end{equation}
As a result, provided that the channel path information $ \mathbf{z}[t] $ is known, the channel matrix $ \mathbf{H}[t] $ can be reconstructed based on (7). It is worth noting that there have been existing works that propose to estimate $ \mathbf{z}[t] $ first before estimating the MIMO channel matrix \cite{tutora}. In contrast, with the concept of CPM, instead of estimating $ \mathbf{z}[t] $ in real time relying on sophisticated training, which incurs non-negligible overhead, it can be inferred merely based on the UE location information, as elaborated in the following. 
With the BS location fixed, the variation of the channel path information $ \mathbf{z}[t] $ is mainly due to the change of UE locations $\mathbf {q}[t]$, and/or the change of environment $E[t]$. Thus, $ \mathbf{z}[t] $ can be expressed in an abstract form as
\begin{equation}
	\mathbf{z}[t]=g_2(\mathbf{q}[t], E[t]),  \ t=1,2,....,T.
\end{equation}
By combining (7) and (9), we have simplified the channel model in (6) into a deterministic function in (7) based on a new model in (9) that has much lower dimensions than that in (6). 
Therefore, as long as a proper CPM is available based on (9), i.e, mapping the UE location $ \mathbf{q}[t]\in \mathcal{Q} $ to the path information $ \mathbf{z}[t] $, the MIMO channel matrix can be reconstructed at each time instant, based on which beam selection according to (4) can be performed. The main procedure of CPM-enabled beam alignment is shown in Fig. 3(a).

%\begin{equation} 		 			 		
%	(\hat{\mathbf{f}}[t], \hat{\mathbf{w}}[t])=\mathop{\arg \max}_{\mathbf{f}_{i} \in \mathcal{F},\mathbf{w}_{j} \in \mathcal{W}}|\mathbf{w}_{j}^\mathrm{H}\hat{\mathbf{H}}[t]\mathbf{f}_{i} |^2,  	 		 	
%\end{equation}

%The process of the CPM enabled training-free beam alignment is shown in the Algorithm 1.
%
%\begin{algorithm}[h]
%	\caption{CPM Enabled Training-Free Beam Alignment} % 名称
%	\label{alg::known CSI}
%	\begin{algorithmic}[1]
%		\Require
%		BS and UE have $ \mathcal{F} , \mathcal{W} $, the UE location
%		\Ensure
%		Index $\hat{i}$ , $\hat{j}$ of estimated beam pair $ \mathbf{\hat{f}} $,  $ \mathbf{\hat{w}} $
%		\State initialize $i=1$ and $j=1$ 
%		\State BS use the UE location to obtain the path information $\mathbf{z}$ from the CPM
%		Channel reconstruction as (5)
%		\For {$i =1,\dots, M_t$} 
%		\For {$j =1,\dots, M_r$}
%		\State The BS calculate $ g_{i,j}=\left|\mathbf{w}_{j}^\mathrm{H} \mathbf{\hat{H}}[t] \mathbf{f}_{i}\right|^2$
%		\EndFor
%		\EndFor
%		\State $\mathbf{g}=\left[ g_{1,1},g_{1,2},\ldots g_{1,M_r}, g_{2,M_r},\ldots,g_{M_t,M_r}\right] $
%		\State$ \hat{i}$ , $\hat{j}=\mathop{\arg \max}\limits_{1\leq i\leq M_t,1\leq j\leq M_r}\mathbf{g}$
%	\end{algorithmic}
%\end{algorithm}

Note that different from the training-based beam alignment, the CPM-enabled beam alignment only requires the UE location information, which requires no extra training overhead and can be readily obtained in contemporary wireless systems. Thus, the effective communication rate becomes
\begin{equation} 		
	\begin{aligned}			
		R&=\frac{1}{T}\sum\limits_{t=1}^{T}  \log \left(1+\frac{P\left|\hat{\mathbf{w}}^{H}[t] \mathbf{H}[t] \hat{\mathbf{f}}[t]\right|^{2}}{\sigma^{2}}\right) \\
		&\leq R^*,
	\end{aligned} 	
\end{equation}
where $ \hat{ \mathbf{w}}[t] $ and $ \hat {\mathbf{f}}[t] $ are the selected beam pairs based on the reconstructed MIMO matrix with CPM and UE location, and the inequality is due to the discrepancy between the reconstructed channel and the true channel in general. By comparing (5) and (10), it is observed that the CPM-enabled training-free beam alignment is able to eliminate the pre-log factor in (5) and thus approach $ R^* $ more closely. Furthermore, provided that UE location and CPM are sufficiently accurate, this rate improvement is more evident with larger $ M_rM_t $. This makes the CPM highly appealing for future wireless systems with extremely large array size and progressively improved localization accuracy.  

\subsection{BIM Enabled Training-Free Beam Alignment}
%In the previous section, we propose the scheme that employs the CPM to store channel path information and reconstruction the channel based on  the information predicted without training overhead. 
%However, there are some the limitations of the CPM: i) Difficulty in obtaining the path parameters. ii) Storage of continuous values. iii) Additional exhaustive beam search with the reconstructed channel.
Note that the CPM-based approach still needs to reconstruct the MIMO channel for beam alignment, which motivates our second proposed CKM instance to directly provide the optimal beam index based on the UE location, which thus called BIM. Specifically, BIM aims to learn the index of the optimal beam pair for all potential UE locations within the BS coverage.
This is motivated by the fact that the ultimate purpose of beam alignment is to select the best beam pair, where channel reconstruction just serves an intermediate step. Furthermore, by combining (3) and (6), the optimal beam pair can be expressed as an abstract function of the UE location $ \mathbf{q}[t] $ and environment $ E[t] $ as
\begin{equation}
	(\mathbf{w}_\text{opt}[t], \mathbf{f}_\text{opt}[t])=g_3(\mathbf{q}[t], E[t]), \ t=1,2,....,T.
\end{equation}
As a result, provided that a sufficiently accurate approximation is made to $ g_3(.,.) $ in (11), i.e., mapping all potential UE locations $ \mathbf{q}[t]\in \mathcal{Q}$ to the corresponding optimal beam pairs, the environment-aware and training-free beam indices can be directly obtained with the UE location information. Furthermore, compared to the channel-reconstruction based approach, the BIM-based approach can avoid the errors and computation overhead due to the channel reconstruction and exhaustive beam search, respectively.
A similar expression for the effective communication rate and its upper bound as (10) can be obtained for the BIM-enabled training-free beam alignment. The main process of BIM-enabled beam alignment is shown in Fig. 3(b).

It is observed from Fig. 3 that compared with CPM-enabled beam alignment, the BIM-based counterpart not only simplifies the process, but also reduces the required storage space (since finite number of beam indices rather than continuous path information needs to be obtained) and the computation overhead (since beam selection as in (4) is avoided). On the other hand, comparing to BIM, the main advantages of CPM lies in that it offers the intrinsic radio propagation channel characteristics, and is independent of the array configuration nor the beamforming codebook. Thus, the choice between CPM and BIM depends on the practical application requirements.  

%\begin{figure}[htbp] 				\centering{\includegraphics[width=.4\textwidth]{CPMBIM.jpg}}  	 		 		\caption{Procedure comparison of CPM- versus BIM-enabled beam alignment.} \label{fig}  	 		 	
%\end{figure}
\begin{figure}[htbp]
	\centering
	\subfigure[CPM]{
		\begin{minipage}[t]{0.25\textwidth}
			\centering
			\includegraphics[width=1\textwidth]{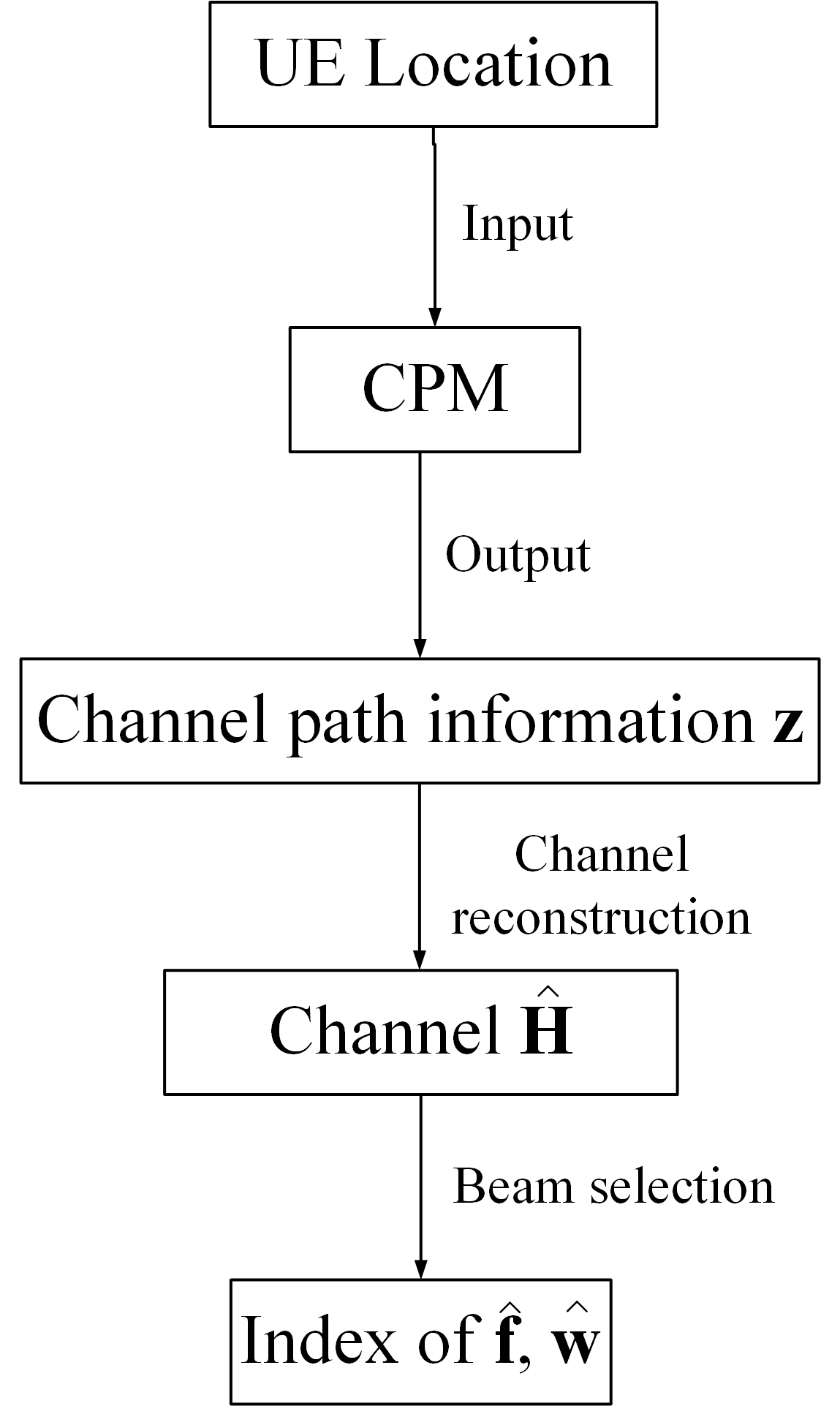}
			%\caption{fig1}
		\end{minipage}%
	}%
	\subfigure[BIM]{
		\begin{minipage}[t]{0.25\textwidth}
			\centering
			\includegraphics[width=0.6\textwidth]{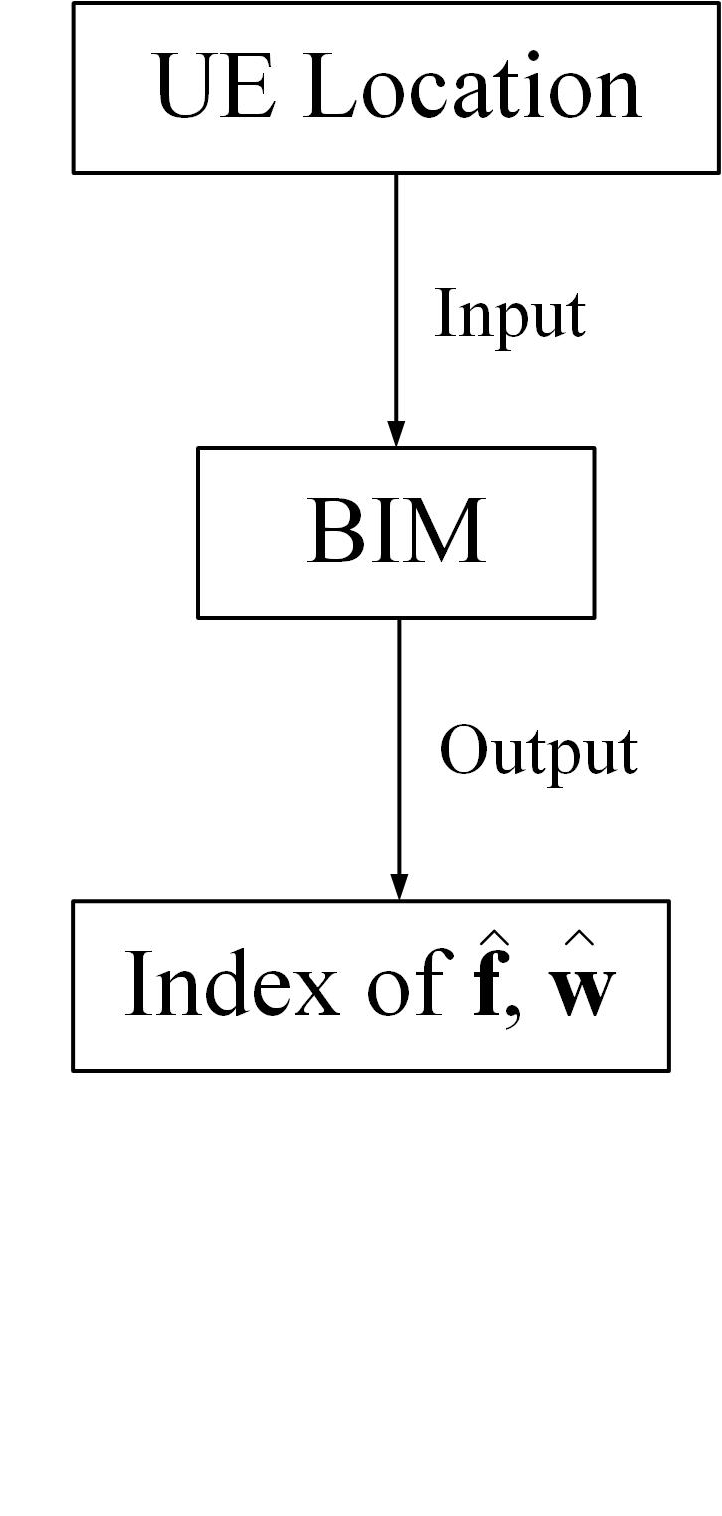}
			%\caption{fig2}
		\end{minipage}%
	}%
	\centering
	\caption{Procedure of CPM- versus BIM-enabled beam alignment.}
\end{figure}

\subsection{Construction of CPM and BIM}	
The performance of the proposed CPM- and BIM-enabled beam alignment techniques critically depends on the accuracy of the corresponding map construction and the estimated UE location. While the latter is expected to be continuously improved with the advancement of a multitude of localization technologies, the former depends on the amount and the quality of the data available to learn the maps. As discussed in \cite{CKM}, the data could be acquired via offline numerical computations based on ray tracing or online/offline data collection. With offline computations, location-specific channel knowledge could be generated using the available physical environment information (such as 3D city or terrain map) and their radio properties (such as dielectric parameters). With offline measurements, dedicated measurement devices such as automatic ground vehicles could be dispatched offline for data collection. 
The data could also be collected online when actual communication takes place. Such offline/online measurement and simulation methods could be jointly applied to augment data for better accuracy.
Based on such finite data points collected, various machine learning techniques such as K-nearest neighbor (KNN)\cite{KNN}, Kridging-based method\cite{Kridge}, or Deep Neural Networks (DNNs)\cite{DNN} could be applied to train the entire CPM or BIM.

\section{Simulation Result}
We consider an actual physical environment shown in Fig. 4, where one BS with location labelled as `Tx' is shown. The commercial ray tracing software Remcom Wireless Insite\footnote{https://www.remcom.com/wireless-insite-em-propagation-software} is used to generate the ground-truth channel information, including the power, phase, AoA, AoD of the channel paths at each location. Two example receiver locations are also shown in the figure, together with their three strongest channel paths. 
The BS is assumed to be equipped with $ M_t^z \times M_t^y $ uniform planar array (UPA) with adjacent elements separated by half wavelength, where the $ M_t^z $ is fixed  to 10 and $ M_t^y $ varies from 5 to 150. Therefore, the total number of BS antennas $ M_t $ varies from 50 to 1500. The Kronecker product based beamforming codebook \cite{codebook} is used, with
		\begin{equation}
		\begin{array}{l}
			\mathbf{c}_{m}^{y}=\frac{1}{\sqrt{M_t^{y}}}\left[1, e^{\frac{j 2 \pi m}{M_t^y}}, \ldots, e^{\frac{j 2 \pi\left(M_t^{y}-1\right) m}{M_t^y}}\right]^{T} ,\\
			\mathbf{c}_{n}^{z}=\frac{1}{\sqrt{M_t^{z}}}\left[1, e^{\frac{j 2 \pi n}{ M_t^z}}, \ldots, e^{\frac{j 2 \pi\left(M^{z}_t-1\right) n}{ M^z_t}}\right]^{T}, \\
			\mathbf{c}_{m,n}=\mathbf{c}_{m}^{y} \otimes \mathbf{c}_{n}^{z},
		\end{array}
		\end{equation}
where {{\small$ m=0,1, \ldots, M^{y}_t-1, n=0,1, \ldots, M^{z}_t-1$}}, while {{\small$ \otimes $}} represents the Kronecker product.
Furthermore, the UE is assumed to be equipped with $5\times5$ UPA, with a similar Kronecker product based beamforming codebook as (12). 
\begin{figure}[htbp] 
	\centering{\includegraphics[width=.3\textwidth]{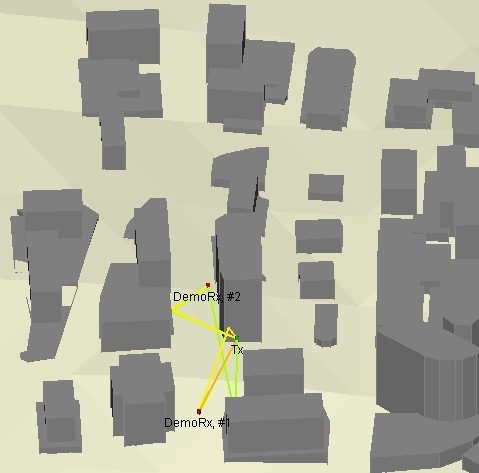}}  	 	
	\caption{ Physical environment for numerical simulations\cite{CKM}.} \label{fig}  	 		 	 		
\end{figure}

A CPM is constructed to learn the $ L=3 $ strongest channel paths, which gives their corresponding path gain, phase, zenith and azimuth AoDs and AoAs. 
Based on the finite set of 38308 ray tracing data samples, the inverse distance weighting (IDW) method of the K-nearest neighbors (with $ K=3 $) is used to construct the entire CPM of all locations.
Furthermore, a BIM is constructed to learn the optimal beam pairs at each location. The training data are obtained with the optimally selected beam pairs for a finite set of 38308 randomly selected UE locations and their corresponding channel using ray tracing simulation. The KNN method with $ K=3 $ is used to learn the entire BIM based on such finite training samples.	
	
For the benchmark schemes, besides the beam sweeping based beam alignment, we also consider another training-free technique, namely the location-based beam alignment, as well as the ideal case based on perfect CSI. For location-based  beam alignment\cite{location}, the AoAs/AoDs are simply calculated based on the relative positions of the BS and UE, but irrespective of the propagation environment.
The channel coherent time is assumed to span over $ N=5 \times 10^4 $ symbols, and all simulations given below are averaged over 1000 randomly selected UE locations. 
	
Fig. 5 shows the average effective communication rate versus the number of transmit antennas $ M_t $ for various beam alignment schemes. For the location-based and CKM-based schemes, the UE location is assumed to be perfectly known without any error. 
As shown in Fig. 5, the effective communication rate of the beam sweeping schemes begins to decrease drastically after the number of BS antennas $ M_t $ grows to 500, which is due to that the training overhead outweighs the resulting beamforming gain for large antenna systems. Furthermore, it is observed that the location-based scheme leads to poor performance for all $ M_t $ considered, since it is ignorant of the actual propagation environment. In contrast, the proposed CPM- and BIM-enabled beam alignment schemes significantly outperform the location-based and beam-sweeping schemes, thanks to its significant saving of training overhead, together with the environment-awareness enabled by CKM. Besides, both proposed schemes give close performance to that based on perfect CSI, e.g., around 90\% when the number of transmit antennas is 1000. 

It is further observed from Fig. 5 that the BIM-based scheme achieves slightly better performance than CPM-based scheme for moderate $ M_t $, due to its end-to-end overall learning capability without relying on the intermediate channel reconstruction, which thus avoids error propagation. With the further increase of the number of antennas $ M_t $, the performance of the CPM-based scheme and BIM-based scheme converge gradually, due to the extremely sharp beam formed. 
	\begin{figure}[htbp]
	\centering{\includegraphics[width=.5\textwidth]{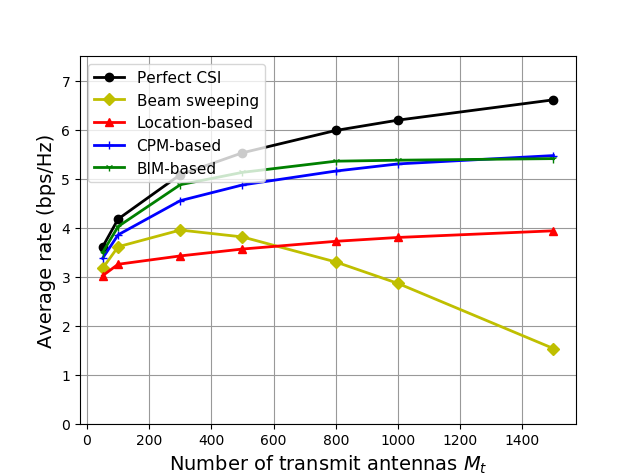}}  	 	
	\caption{Comparison of average effective communication rate for various beam alignment schemes. The UE location is assumed to be perfectly known.} \label{fig}  	 		 	 		
\end{figure}

To study the impact of UE location error on the performance of the proposed schemes, Fig. 6 plots the average effective communication rate for various schemes with average location error of 1 meter, which is modeled based on Rayleigh distribution. 
By comparing Fig. 5 and Fig. 6, it is observed that the location error only slightly degrades the rate performance of the proposed CPM-based and BIM-based schemes. Furthermore, even with such a moderate location error, these two proposed schemes still significantly outperform the benchmark beam sweeping and location-based schemes. This is expected since while the UE location error slightly degrades the resulting beamforming gain that only affects the rate logarithmically, the saving of the training overhead that affects the communication rate linearly is still dominating. This demonstrates the great potential of the proposed CKM-enabled beam alignment for large-scale MIMO systems. 
	\begin{figure}[htbp]
	\centering{\includegraphics[width=.5\textwidth]{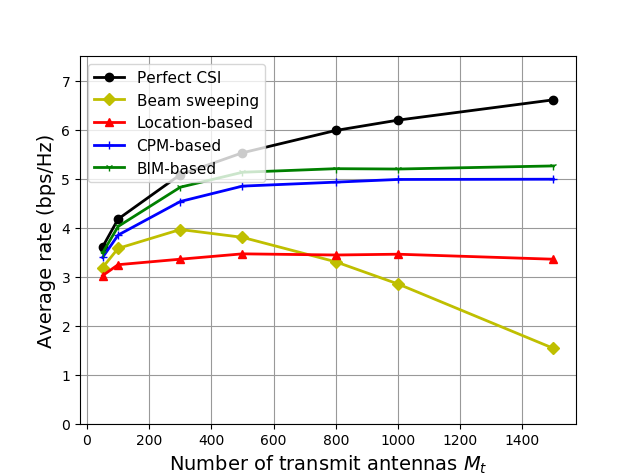}}  	 		 		\caption{Comparison of average effective communication rate for various beam alignment schemes. The UE location is assumed to have average location error of 1 meter.} \label{fig}  	 		 	 		 	
\end{figure}

\section{Conclution}
In this paper, a novel environment-aware beam alignment technique enabled by CKM is proposed, which significantly saves the prohibitive training overhead of conventional training-based schemes. Two specific instances of CKM are introduced, namely CPM- and BIM-enabled beam alignment. Simulation results over realistic radio propagation environment based on ray-tracing demonstrate that the proposed schemes significantly outperform the benchmark beam sweeping and location-based schemes, and even achieve rate performance very close to the ideal case with perfect CSI.  

\section*{Acknowledgment}
This work was supported by the National Key R\&D Program of China with Grant number 2019YFB1803400.

\bibliographystyle{IEEEtran}
\bibliography{ref2}
\end{document}